\documentclass[prb, reprint, twocolumn, superscriptaddresss, longbibliography]{revtex4-2}
\def\be{\begin{equation}}
\def\ee{\end{equation}}
\def\bea{\begin{eqnarray}}
\def\eea{\end{eqnarray}}
\def\bi{\begin{itemize}}
\def\ei{\end{itemize}}
\usepackage{braket}

\usepackage{xcolor}
\usepackage{graphicx}
\usepackage{amsmath,amstext,amssymb,bm,color,times}
\usepackage[english]{babel}
\usepackage[T1]{fontenc}
\usepackage[utf8]{inputenc}
\usepackage[colorlinks=true,citecolor=blue,linkcolor=magenta]{hyperref}
\usepackage{lipsum}
\usepackage{adjustbox}

\begin{document}

\title{Kink correlations, domain size distribution, and the emptiness formation probability \\
           after the Kibble-Zurek quench in the quantum Ising chain} 

\author{Jacek Dziarmaga}
\author{Marek M. Rams}
\affiliation{Jagiellonian University, Institute of Theoretical Physics, 
             ul. {\L}ojasiewicza 11, PL-30348 Krak\'ow, Poland}
\date{\today}

\begin{abstract}
Linear quench of the transverse field drives the quantum Ising chain across a quantum critical point from the paramagnetic to the ferromagnetic phase. We focus on normal and anomalous quadratic correlators between fermionic kink creation and annihilation operators. They depend not only on the Kibble-Zurek (KZ) correlation length but also on a dephasing length scale, which differs from the KZ length by a logarithmic correction. Additional slowing down of the ramp in the ferromagnetic phase further increases the dephasing length and suppresses the anomalous correlator.
The quadratic correlators enter Pfaffians that yield experimentally relevant kink correlation functions, the probability distribution of ferromagnetic domain sizes, and, closely related, emptiness formation probability. The latter takes the form of a Pfaffian of a block Toeplitz matrix that allows for some analytic asymptotes.
Finally, we obtain further insight into the structure of the state at the end of the ramp by interpreting it as a paired state of fermionic kinks characterized by its pair wave function.
All those quantities are sensitive to quantum coherence between eigenstates with different numbers of kinks, thus making them a convenient probe of the quantumness of a quantum simulator platform.

\end{abstract}
\maketitle

\section{ Introduction } 

The original Kibble scenario for defect formation in cosmological symmetry-breaking phase transitions~\cite{K-a, *K-b, *K-c} inspired Zurek mechanism for the dynamics of laboratory phase transitions~\cite{Z-a, *Z-b, *Z-c, Z-d}. It uses equilibrium critical exponents and the transition (quench) time to predict the scaling of the resulting density of topological defects. Kibble-Zurek mechanism (KZM) was supported by numerical simulations~\cite{KZnum-a,KZnum-b,KZnum-c,KZnum-d,KZnum-e,KZnum-f,KZnum-g,*KZnum-h,*KZnum-i,KZnum-j,KZnum-k,KZnum-l,KZnum-m,that} and experiments in condensed matter systems~\cite{KZexp-a,KZexp-b,KZexp-c,KZexp-d,KZexp-e,KZexp-f,KZexp-g,KZexp-gg,KZexp-h,KZexp-i,KZexp-j,KZexp-k,KZexp-l,KZexp-m,KZexp-n,KZexp-o,KZexp-p,KZexp-q,KZexp-r,KZexp-s,KZexp-t,KZexp-u,KZexp-v,KZexp-w,KZexp-x}. More recently, it was generalized to quantum phase transitions~\cite{QKZ1,QKZ2,QKZ3,d2005,d2010-a,d2010-b}, with an outbreak of theoretical developments~\cite{QKZteor-a,QKZteor-b,QKZteor-c,QKZteor-d,QKZteor-e,QKZteor-f,QKZteor-g,QKZteor-h,QKZteor-i,QKZteor-j,QKZteor-k,QKZteor-l,QKZteor-m,QKZteor-n,QKZteor-o,KZLR1,KZLR2,KZLR3,QKZteor-q,QKZteor-r,QKZteor-s,QKZteor-t,sonic,QKZteor-u,QKZteor-v,QKZteor-w,QKZteor-x,RadekNowak,intereferometryChRL,assessingGardas} as well as a series of experimental tests~\cite{QKZexp-a, QKZexp-b, QKZexp-c, QKZexp-d, QKZexp-e, QKZexp-f, QKZexp-g,deMarco2,Lukin18,2dkzdwave,adolfodwave,King_Dwave1d_2022} of the quantum KZM (QKZM). Notably, the emulation of the transverse field quantum Ising chain with Rydberg atoms~\cite{Lukin18} is consistent with the theoretically predicted scalings~\cite{QKZ2,QKZ3,d2005}. Furthermore, the recent quantum annealing experiment with the {\it coherent D-Wave}~\cite{King_Dwave1d_2022}, emulating the same Ising chain, goes beyond these scaling predictions probing the probability distribution for the number of kinks~\cite{QKZteor-e,delcampostatistics} and kink-kink correlations in space~\cite{roychowdhury2020dynamics,RadekNowak}. This experiment not only supports QKZM but, by no means less important, makes quantumness of the new annealer more likely, in contrast to its older incarnations, where it was questionable~\cite{QKZexp-g,2dkzdwave,adolfodwave}.

In this article, we aim at providing further experimentally relevant characteristics of the post-quench state in the transverse field Ising chain. These include distribution of the domain sizes, the emptiness formation probability, and higher kink-kink correlators. 

The most simple picture of QKZM assumes an adiabatic-impulse-adiabatic approximation, where the evolution at a distance from the critical point is adiabatic, and the state of the system freezes out in the neighborhood of the critical point due to the closing of the energy gap. While such an approximation has deficits, it predicts the correct KZM scaling laws. A more detailed causality-based picture, emphasizing the role of the sonic horizon and the speed of the relevant sound, is discussed in Ref.~\onlinecite{sonic}.

In QKZM scenario a system, initially prepared in its ground state, is smoothly ramped across a continuous quantum critical point. A generic ramp can be linearized near the critical point as
\begin{equation}
\epsilon(t)=\frac{t-t_c}{\tau_Q}.
\label{epsilont}
\end{equation}
Here $\epsilon$ is a dimensionless parameter in the Hamiltonian, whose magnitude measures the distance from the critical point, $\tau_Q$ is the quench time, and $t_c$ is the instance of a time when the critical point is crossed. Initially, the evolution is adiabatic and the system follows its instantaneous ground state. The adiabaticity fails at $t_c-\hat t$ when the reaction rate of the system, proportional to its energy gap $\Delta\propto|\epsilon|^{z\nu}$, equals instantaneous transition rate, proportional to $|\dot \epsilon/\epsilon| = 1/|t-t_c|$. Here $z$ and $\nu$ are, respectively, the dynamical and the correlation length critical exponent. The equality implies
\be 
\hat t\propto \tau_Q^{z\nu/(1+z\nu)},
\label{hatt}
\ee 
and the corresponding $\hat\epsilon=\hat t/\tau_Q\propto\tau_Q^{-1/(1+z\nu)}$. In the cartoon impulse approximation, the ground state at $-\hat\epsilon$, with a corresponding correlation length,
\begin{equation}
\hat\xi \propto \hat\epsilon^{-\nu} \propto \tau_Q^{\nu/(1+z\nu)},  
\label{hatxi}
\end{equation}
is expected to survive until $t_c+\hat t$, when the adiabatic evolution can restart [in a more complete sonic horizon argument, the correlation length gets increased by a factor of $\mathcal{O}(5)$]. In this way, $\hat\xi$ becomes imprinted on the initial state for the final adiabatic stage of the evolution after $+\hat t$. The adiabatic-impulse-adiabatic approximation predicts the correct scaling of the characteristic length scale $\hat \xi$ with $\tau_Q$, see Eq.~\eqref{hatxi}, and the timescale $\hat t \propto \hat\xi^z$. The post-quench density of excitations is determined by $\hat\xi$ within this scenario.  

In the integrable quantum Ising chain, the excitations are well defined as Bogoliubov quasiparticles~\cite{d2005}. They get excited between $t_c-\hat t$ and $t_c+\hat t$, and later their power spectrum $p_k$ remains frozen when the evolution becomes adiabatic again after $t_c+\hat t$. Here $p_k$ is the excitation probability for a pair of quasiparticles with quasimomenta $\pm k$. The state after $t_c+\hat t$ is a quantum superposition over many eigenstates with amplitudes that have frozen magnitudes (determined by $p_k$), which, however, accumulate dynamical $k$-dependent phases. When given enough time, this may eventually lead to dephasing: the phases become scrambled enough to allow accurate calculation of local observables  within a random phase approximation~\cite{QKZteor-e}.

While the average density of kinks~\cite{d2005}---and even their number distribution~\cite{QKZteor-e,delcampostatistics,adolfodwave}--- depend only on the power spectrum $p_k$, the kink-kink correlation function is sensitive to the dynamical phase~\cite{roychowdhury2020dynamics,RadekNowak}. At first, this correlator was obtained in the random phase approximation~\cite{roychowdhury2020dynamics}. In the follow-up paper~\cite{RadekNowak}, an analytic formula was derived that includes an extra term reflecting phase coherences. It has a characteristic peak that served as one of the hallmarks of quantumness in the {\it coherent D-Wave} experiment~\cite{King_Dwave1d_2022}. The present paper simplifies the formalism of Ref.~\onlinecite{RadekNowak} by employing fermionic kink creation/annihilation operators. This formulation facilitates the calculation of higher-order kink correlators at the end of the KZ quench as well as the domain size distribution and the emptiness formation probability. 

The paper is organized as follows. In Sec.~\ref{sec:QIM} we outline the basics of the quantum Ising chain. In Sec.~\ref{sec:kinks} the fermionic kink annihilation operators enter the stage. In Sec.~\ref{sec:linearquench} a linear KZ ramp of the transverse field is mapped to the Landau-Zener model to obtain the excitation probability $p_k$, and the dynamical phases $\varphi_k$. Quadratic fermionic kink correlators characterizing the Gaussian state after the quench are derived in Sec.~\ref{sec:quadratic}. In Sec.~\ref{sec:KKcorrelators} the $M$-kink correlation functions are expressed as Pfaffians of skew-symmetric $2M\times2M$ matrices made of the quadratic correlators. A special type of a multi-kink correlator is the domain size distribution constructed in Sec.~\ref{sec:KKdomains}. In Sec.~\ref{sec:EL} the emptiness formation probability is expressed by a Toeplitz matrix, and its relation to the domain size distribution is established. This Toeplitz property allows for some analytic results collected in Sec.~\ref{sec:analytic}. In Sec.~\ref{sec:Cooper} a pair wave function is derived that characterizes the paired state of kinks excited by the quench. We conclude in Sec.~\ref{sec:summary}. The two appendices include, in App.~\ref{app:integral}, the technical details of the derivation of the anomalous fermionic kink correlator and,   in App.~\ref{app:slowing}, comments on the accumulation of extra dephasing during a non-linear ramp that slows down in the ferromagnetic phase.

\section{ Quantum Ising chain }
\label{sec:QIM}

The transverse field quantum Ising chain is
\be
H~=~-\sum_{n=1}^N \left( g\sigma^x_n + \sigma^z_n\sigma^z_{n+1} \right),
\label{Hsigma}
\ee
where we assume periodic boundary conditions, $\vec\sigma_{N+1}~=~\vec\sigma_1$, and even $N$ for definiteness.
In the thermodynamic limit, $N\to\infty$, the quantum critical point at $g=1$ separates the paramagnetic ($|g|>1$) from the ferromagnetic ($|g|$<1) phases.  After the Jordan-Wigner transformation:
\bea
&&
\sigma^x_n~=~1-2 c^\dagger_n  c_n~, \nonumber \\
&&
\sigma^z_n~=~
-\left( c_n^\dagger + c_n \right)
 \prod_{m<n}(1-2 c^\dagger_m c_m)~,
 \label{JW}
\eea
introducing fermionic annihilation operators $c_n$, the Hamiltonian (\ref{Hsigma}) becomes
\be
 H~=~P^+~H^+~P^+~+~P^-~H^-~P^-~.
\label{Hc}
\ee
Here the projectors
\be
P^{\pm}=
\frac12\left[1\pm\prod_{n=1}^N\sigma^x_n\right]=
\frac12\left[1~\pm~\prod_{n=1}^N\left(1-2c_n^\dagger c_n\right)\right],
\label{Ppm}
\ee
define subspaces with even/odd numbers of fermions as the parity, related to $\prod_n \sigma^x_n$, is a good quantum number.
\bea
H^{\pm}~=~
\sum_{n=1}^N
\left[ 
g \left(c_n^\dagger  c_n -\frac12 \right) - c_n^\dagger  c_{n+1} + c_{n}  c_{n+1} \right] + {\rm h.c.} ~
\label{Hpm}
\eea
are the corresponding reduced Hamiltonians. Fermionic operators $c_n$ in $H^\pm$ satisfy (anti-)periodic boundary conditions: $c_{N+1}=\mp c_1$. 

The ground state has even parity for any non-zero $g$. For a time evolution that begins in the ground state, the state remains in the even parity subspace. Relevant $H^+$ can be simplified by an anti-periodic Fourier transform: 
\be
c_n~=~ 
\frac{e^{-i\pi/4}}{\sqrt{N}}
\sum_k c_k e^{ikn}~,
\label{Fourier}
\ee
where pseudomomenta take half-integer values:
\be
k~=~
\pm \frac12 \frac{2\pi}{N},
\dots,
\pm \frac{N-1}{2} \frac{2\pi}{N}~.
\label{k}
\ee
It brings the Hamiltonian to an additive form,
\be 
H^+ = \sum_{k>0} H_k, \label{Hck}
\ee 
where 
\bea
H_k &=&
2 (g-\cos k) \left(c_k^\dagger c_k - c_{-k} c_{-k}^\dagger \right)+  \nonumber\\
&&
\ \ \ \ \ \ \ \ \ \
2 \sin k
\left( c^\dagger_k c^\dagger_{-k} + c_{-k} c_k \right).
\label{Hk}
\eea
Its diagonalization could be completed by a Bogoliubov transformation in $\pm k$ momentum subspace, but here we prefer to take a more pedestrian approach. 

Each Hamiltonian $H_k$ lives in a 4-dimensional subspace but its ground state, and also the state during the time ramp, belongs to a 2-dimensional subspace spanned by
\be 
\ket{\psi_k} = u_k^*~ \ket{0_k} + v_k^*~ c_{-k}^\dag c_k^\dag \ket{0_k},
\label{uv}
\ee 
where $\ket{0_k}$ is a state without fermions: $c_{\pm k}\ket{0_k}=0$. The eigenstates satisfy stationary Bogoliubov-de Gennes (BdG) equations:
\bea
\varepsilon 
\left[ 
\begin{array}{c}
u_k \\
v_k 
\end{array}
\right] =
2
\left[
\begin{array}{cc}
g-\cos k & \sin k  \\
\sin k     & -g + \cos k 
\end{array}
\right] 
\left[
\begin{array}{c}
u_k \\
v_k 
\end{array}
\right], 
\label{stBdG}
\eea
where eigenvalues $\varepsilon$ are minus eigenenergies of Hamiltonians in Eq.~\eqref{Hk}. We assume the complex conjugations in the ansatz~\eqref{uv}, $u^*_k$ and $v^*_k$,  for the Bogoliubov modes $(u_k,v_k)$ to satisfy the same equations as in Ref.~\cite{d2005,RadekNowak}. There, the same problem was treated by the Bogoliubov formalism without writing down the explicit wave function in Eq.~\eqref{uv}. 

There are two eigenenergies corresponding to $\varepsilon=\pm\varepsilon_k$, where  
\be
\varepsilon_k~=~2\sqrt{(g-\cos k)^2+\sin^2 k}~.
\label{epsilonk}
\ee
This is a quasiparticle dispersion~\cite{d2005,RadekNowak}. At critical $g=1$, the dispersion is linear for small $|k|$, $\varepsilon_k\approx 2|k|$, that implies the dynamical exponent $z=1$. Furthermore, for $k=0$, we have $\varepsilon_0\propto |g-1|^1$ when $g\approx1$, that implies $z\nu=1$ and the correlation length exponent $\nu=1$. Accordingly, the KZ scales are
\be 
\hat t \propto \hat\xi \propto \sqrt{\tau_Q},
\label{KZising}
\ee 
up to a numerical prefactors ${\cal O}(1)$.

\section{ Fermionic kink quasiparticles }
\label{sec:kinks}

We introduce a {\it fermionic} kink operator that annihilates a kink on a bond connecting sites $n$ and $n+1$:
\be 
\gamma_{n +\frac12} \equiv  \left( \prod_{l\le n} \sigma^x_l\right) \frac{ \sigma^z_{n} - \sigma^z_{n+1}}{2 i}.
\label{gamma}
\ee
Here the $\sigma^x$-string flips all the spins to the left of the bond. Before the flip, the $\sigma^z$-factor checks if there is a kink on the bond and at the same time changes the parity.
A phase factor is included for later convenience. Overall, the operator annihilates the kink, if there is one, or gives zero otherwise. For our periodic boundary conditions this operator seemingly has an unwanted effect on the bond connecting sites $N$ and $1$, however, in the following, we consider only products of even numbers of kink creation/annihilation operators where these effects cancel out. With the Jordan-Wigner transformation in Eq.~\eqref{JW}, it translates to
\bea 
\gamma_{n +\frac12} =
\frac{1}{2i}
\left( 
c_{n+1}-c_{n}+c_{n+1}^\dag+c^\dag_{n}
\right),
\label{gamman}
\eea  
and its Fourier transform is
\bea 
\gamma_k & = & 
\frac{ e^{i\pi/4} }{ \sqrt{N} }
\sum_n \gamma_{n+\frac12} e^{-ik(n+\frac12)} \nonumber\\
& = &
c_k \sin\frac{k}{2} + c_{-k}^\dag \cos \frac{k}{2} .
\label{gammak}
\eea 
This formula can be understood as a Bogoliubov transformation.

In the ferromagnetic Ising chain at $g=0$, Eq.~\eqref{stBdG} has two eigenstates with eigenvalues $\varepsilon=2$ and $\varepsilon=-2$, respectively,
\bea 
\ket{\rm GS} &=&
\sin\frac{k}{2} \ket{0_k} + \cos\frac{k}{2} c_{-k}^\dag c_k^\dag \ket{0_k},
\label{GS} \\
\ket{\rm ES} &=&
\cos\frac{k}{2} \ket{0_k} - \sin\frac{k}{2} c_{-k}^\dag c_k^\dag \ket{0_k} . 
\label{ES}
\eea 
The ground state is a no-kink vacuum, $\gamma_{\pm k}\ket{\rm GS}=0$, and the excited state consists of a pair of kinks:
$\ket{\rm ES}=\gamma_{-k}^\dag \gamma_k^\dag \ket{\rm GS}$. It follows that diagonalized Hamiltonian at $g=0$ reads
\be 
H^+ = E_0 + 2 \sum_k \gamma_k^\dag \gamma_k, 
\ee 
where $E_0$ is the vacuum energy. Due to the projector $P^+$ in Eq.~\eqref{Hc}, only states with even numbers of kinks belong to the spectrum of $H^+$.

\section{ Linear quench }
\label{sec:linearquench}

We ramp the system across the quantum critical point as
\be
g(t\leq0)~=~-\frac{t}{\tau_Q}~.
\label{linear}
\ee
with the characteristic quench time $\tau_Q$. The ramp crosses the critical point at $t_c=-\tau_Q$. For the most universal features of the QKZM, it is enough to assume that the ramp can be linearized near the critical point with a slope equal $-1/\tau_Q$. Here, we will proceed with an analytic solution of the linear ramp. The system is initially in its ground state at the large initial value of $g\gg 1$ but, as $g$ is ramped down to zero, it gets excited from its instantaneous ground state and the final state, at $t=0$, has a finite number (density) of kinks. 

The states in Eq.~\eqref{uv} evolve according to the time-dependent Bogoliubov-de Gennes equations: 
\bea
i\frac{d}{dt}
\left[
\begin{array}{c}
u_k \\
v_k 
\end{array}
\right] =
2
\left[
\begin{array}{cc}
g-\cos k & \sin k  \\
\sin k     & -g + \cos k 
\end{array}
\right]
\left[ 
\begin{array}{c}
u_k \\
v_k 
\end{array}
\right].
\label{dynBdG}
\eea
These equations can be mapped to the Landau-Zener problem and the final state at the end of the ramp reads~\cite{d2005,QKZteor-e}
\be 
\left[ 
\begin{array}{c}
u_k(0) \\
v_k(0) 
\end{array}
\right] =
\left[ 
\begin{array}{c}
\sin\frac{k}{2} \\
\cos\frac{k}{2} 
\end{array}
\right]
\sqrt{1-p_k}+
\left[
\begin{array}{c}
\cos\frac{k}{2} \\
-\sin\frac{k}{2} 
\end{array}
\right]
\sqrt{p_k}
e^{i\varphi_k},
\ee 
up to a global phase.
Here, the Landau-Zener excitation probability
\be 
p_k \simeq e^{ - 2\pi \tau_Q k^2 },
\label{pk}
\ee 
and a dynamical phase~\cite{QKZteor-e,RadekNowak}
\be 
\varphi_k \simeq \frac{\pi}{4} + 2\tau_Q + k^2 \tau_Q \ln \tau_Q,
\label{varphik}
\ee 
where both formulas are accurate when $\tau_Q\gg 1$.  In other words, the final state is
\be 
\ket{\psi_k(0)} = 
\left( 
\sqrt{1-p_k} + 
\sqrt{p_k} 
e^{-i\varphi_k} 
\gamma^\dag_{-k} \gamma^\dag_k 
\right) 
\ket{\rm GS},
\label{psi0}
\ee 
where $k>0$.

An average number of kinks in this state is
$ 
{\cal N}~=~\sum_k~p_k.
$
For large $N$, the sum can be replaced by an integral and the density of kinks becomes
\cite{d2005}:
\be
\rho =
\frac{\cal N}{N}=
\frac{1}{2\pi}\int_{-\pi}^{\pi}dk~p_k \simeq
\frac{1}{2\pi\sqrt{2\tau_Q}}.
\label{scaling}
\ee    
The density scales like $\hat\xi^{-1}\propto\tau_Q^{-1/2}$, in agreement with the QKZM. It is natural to make the definition of $\hat\xi$ precise as 
\be 
\hat\xi \equiv 2\pi\sqrt{2\tau_Q} = \frac{1}{\rho},
\label{hatxiprecise}
\ee 
so that an inverse of it is equal to the final density of kinks. 

\section{ Quadratic fermionic Kink correlators }
\label{sec:quadratic}

The Gaussian state Eq.~\eqref{psi0} is fully characterized by its quadratic fermionic kink correlators, normal:
\bea 
N_{R} 
&=& 
\langle \gamma^\dag_{n+R+\frac12} \gamma_{n+\frac12} \rangle
=
\frac{1}{2\pi}\int_{-\pi}^\pi dk~p_k \cos k R \nonumber \\
&\simeq& 
\hat\xi^{-1} e^{-\pi(R/\hat\xi)^2}   ,
\eea 
and anomalous:
\bea 
\Delta_{R} 
&=& 
\langle \gamma_{n+R+\frac12} \gamma_{n+\frac12} \rangle \nonumber \\
&=&
\frac{1}{2\pi}\int_{-\pi}^\pi dk~\sqrt{(1-p_k)p_k} e^{-i\varphi_k} \mathrm{sgn}(k)\sin k R \nonumber \\
&\simeq& 
c
\frac{R}{\sqrt{\hat\xi l^3}}
e^{ -\frac{3\pi}{2} (R/l)^2 }
e^{-i\phi_R}.
\label{Delta}
\eea 
Here $c\simeq 3.0934$ is a constant, the phase
\bea 
\phi_R 
&=&
\frac14\pi+2\tau_Q-\frac32 {\rm arg}\left(1-\frac{3i\ln\tau_Q}{4\pi}\right) + \nonumber\\
&&-
\frac98(R/l)^2\ln\tau_Q,
\label{phiR}
\eea 
and the correlation range
\be 
l  = 
\hat\xi ~ \sqrt{ 1 + \left(\frac{3\ln\tau_Q}{4\pi}\right)^2 }.
\label{l}
\ee 
For extremely slow quenches, when $\ln\tau_Q\gg4\pi/3$, the range of this correlator becomes much longer than $\hat\xi$. The integral in Eq.~\eqref{Delta} has been worked out in Refs.~\onlinecite{RadekNowak,oscillations}, see also appendix~\ref{app:integral}. Furthermore, when the ramp is not linear after crossing the critical point but instead slows down in the ferromagnetic phase then the resulting extra quasiparticle dephasing can make $l$ arbitrarily long, see appendix~\ref{app:slowing} for more details. In that case, all the following kink-kink correlators become functions of two parameters $\hat\xi$ and $l$, which can be considered independent. 

Correlator $N_R$ depends only on power spectrum $p_k$, which builds up between $t_c-\hat t$ and $t_c+\hat t$, but remains frozen after $t_c+\hat t$. Therefore, as much as a smooth ramp can be considered to be linear between $t_c\mp\hat t$, the final $N_R$ does not depend on the ramp after $t_c+\hat t$. In contrast, $\Delta_R$ does depend on how the ramp continues towards the final $g=0$ though the dynamical phase $\varphi_k$, which depends on the exact path. Slowing the ramp in the ferromagnetic phase before $g=0$ increases the dynamical phase. The slowdown increases dephasing length $l$ and the phase factor $\phi_R$. Longer $l$ suppresses the magnitude of the anomalous correlator in Eq.~\eqref{Delta}, eventually making it negligible. For more details and some estimates, see Appendix~\ref{app:slowing}. 

By their very definitions, $N_R$ and $\Delta_R$ are probing quantum coherence between two classical spin configurations with opposite spin polarization of $R$ consecutive spins. In the case of $N_R$, the configurations have the same number of kinks but with one of the kinks displaced by $R$ sites. In the case of $\Delta_R$, the configurations differ by two kinks separated by $R$ sites. In both cases, the $R$ sites have opposite spin polarization. The quadratic correlators would be tricky to measure as in the spin representation they involve string operators of length $R$, see Eqs.~\eqref{JW} and~\eqref{gamma}. However, they can be probed indirectly through kink correlators and probability distributions that we consider in the following.

However, before we proceed it should be stressed that $N_R$ and $\Delta_R$ are expectation values of operators with support on a finite subsystem. The rest of the system can play the role of its environment. Therefore, even though the whole chain remains in a pure state, its subsystem can undergo decoherence. For instance, if the subsystem is two bonds $n+R+\frac12$ and $n+\frac12$ then $\gamma_{n+R+\frac12} \gamma_{n+\frac12}$ is operating in the Hilbert space of the subsystem with a Fock basis: $\ket{\tilde 0}$ and  $\ket{\tilde 2}=\gamma_{n+R+\frac12}^\dag \gamma_{n+\frac12}^\dag \ket{\tilde 0}$, where $\ket{\tilde 0}$ is the state with no kinks in the subsystem. In this basis $\Delta_R$ is coherence between $\ket{\tilde 0}$ and $\ket{\tilde 2}$. This coherence vanishes when phase $\varphi_k$ in Eq.~\eqref{Delta} depends on $k$ strongly enough to suppress the integral. As we will see, the $k$-dependence increases with time due to non-trivial quasiparticle dispersion (i.e., their mobility). We will call this process dephasing to distinguish it from the decoherence of the whole chain that is absent for the unitary evolution. The dephasing can obscure the coherence of the whole chain when probed through $\Delta_R$ with the finite support, but, as we will see, a typical linear quench does not give the dephasing enough time to become significant. The ramp would have to be deliberately slowed down after crossing the quantum critical point for the dephasing to take effect. 

\section{ Kink correlation functions }
\label{sec:KKcorrelators}

Back in the spin representation, the quadratic fermionic correlators are expectation values of highly non-local string operators and, therefore, are hard to measure. However, they can be probed indirectly through $M$-kink correlators:
\be 
C_{m_1,m_2,\dots,m_M} =
\left\langle 
K_{m_1} K_{m_2} \dots K_{m_M}
\right\rangle.
\ee 
Here,
\be 
K_{n+\frac12}=\gamma^\dag_{n+\frac12} \gamma_{n +\frac12}
\label{K}
\ee
is a projector on a state with a kink on the bond between sites $n$ and $n+1$, and we will use half-integer numbers to index them. It is worth observing here that this projector can be also expressed as $K_{n+\frac12}=\overline\gamma^\dag_{n+\frac12} \overline\gamma_{n+\frac12}$, where $\overline\gamma_{n+\frac12}$ are hard-core bosonic kink annihilation operators, $\overline \gamma_{n +\frac12} = -i  \gamma_{n +\frac12}\sigma^z_{n+1} = \left(\prod_{l\le n} \sigma^x_l\right) \frac{ 1 - \sigma^z_{n} \sigma^z_{n+1} }{ 2} = \left(\prod_{l\le n} \sigma^x_l\right) K_{n+\frac12}$.

\begin{figure}[t!]
\vspace{-0cm}
\includegraphics[width=1\columnwidth,clip=true]{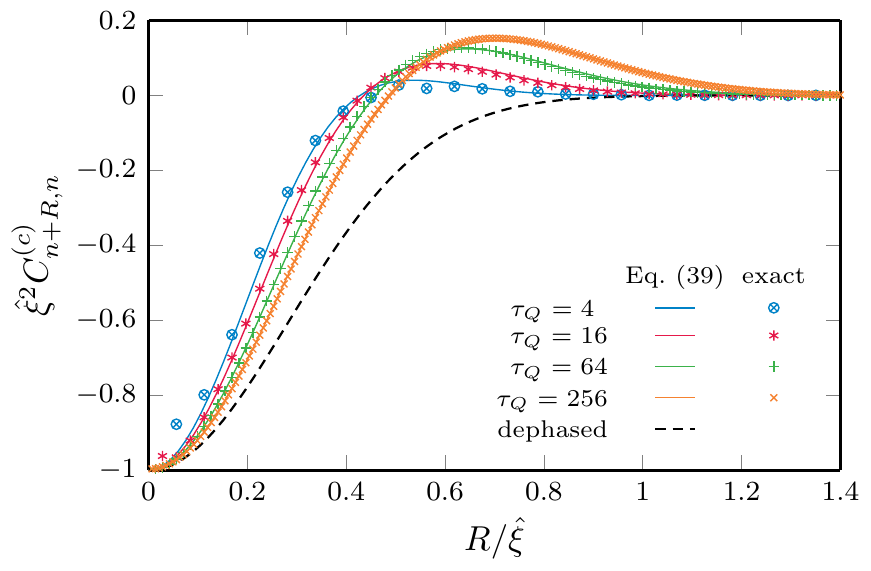}
\vspace{-0cm}
\caption{
{\bf Connected two-kink correlator.}
The solid lines show connected $2$-kink correlators in Eq.~\eqref{Cc} for several values of the quench time $\tau_Q$.
The correlation strength and the range were scaled by the KZ length scale $\hat\xi$. The scaled plots do not collapse because the distribution depends not only on $\hat\xi$ but also on the dephasing length $l$ in Eq.~\eqref{l}.
The dashed line is a scaled correlator for a ramp that slows in the ferromagnetic phase for long enough to inflict a complete quasiparticle dephasing. The dephasing sends $l\to\infty$ making the distribution to depend only on $\hat \xi$.  
Points show the exact results, obtained via numerical integrals of the exact solutions of Eq.~\eqref{dynBdG}. Already for $\tau_Q=4$, they are well reproduced by Eq.~\eqref{Cc}, which is derived under the assumption of $\tau_Q \gg 1$,
except for a few points for small $R$ where this $\tau_Q$ is still not long enough. Even these discrepancies disappear with increasing $\tau_Q$.
}
\label{fig:C2}
\end{figure}

With the help of the Wick theorem we obtain:
\bea 
C_{m} &=& \rho, \\
C_{m+R, m} 
&=& 
\rho^2+ \left| \Delta_{R} \right|^2 - N^2_{R} 
\label{C2} \\
&=&
\rho^2
\left[
1
+
c^2
\left(\hat\xi/l\right)
\left(R/l\right)^2
e^{ -3\pi (R/l)^2 } -
e^{-2\pi(R/\hat\xi)^2}
\right].
\nonumber 
\eea 
A connected part of the $2$-kink correlator is
\bea 
\hat\xi^2 C_{m+R,m}^{(c)} &=& \hat\xi^2 \left( C_{m+R,m} - C_{m+R} C_m \right) \label{Cc} \\
&=&
c^2
\left(\hat\xi/l\right)
\left(R/l\right)^2
e^{ -3\pi (R/l)^2 } -
e^{-2\pi(R/\hat\xi)^2}.
\nonumber
\eea 
It was scaled here by a factor $\hat\xi^2=\rho^{-2}$ to be properly normalized with respect to the average kink density.
The scaled connected correlator is shown in Fig.~\ref{fig:C2}.
This form of the $2$-kink correlator is already known from Ref.~\onlinecite{RadekNowak}, but here it was derived without some unnecessary approximations. When dephasing length $l$ is not too large, the correlator has a peak that served as one of hallmarks of coherent quantum annealing by a D-Wave machine~\cite{King_Dwave1d_2022}, see Fig.~3 in that work. 
With extra dephasing, which would make dephasing length $l$ long enough, the correlator would simplify to 
\be 
\hat\xi^2 C_{m+R, m}^{(c)}=-e^{-2\pi(R/\hat\xi)^2}, 
\ee 
see the dashed plot in Fig.~\ref{fig:C2}.

In the same manner, a 3-kink correlator is
\bea 
&&
C_{m_1,m_2,m_3} = \nonumber\\
&&
\rho^3 +
\left( C_{m_1,m_2} + C_{m_2,m_3} + C_{m_3,m_1}  \right) \rho +
2 {\cal N}_{12} {\cal N}_{23} {\cal N}_{31} + \nonumber\\
&&
2 {\rm Re} 
\left[   
{\cal N}_{12} {\cal D}_{23} {\cal D}^*_{31} +
{\cal N}_{23} {\cal D}_{31} {\cal D}^*_{12} +
{\cal N}_{31} {\cal D}_{12} {\cal D}^*_{23}
\right].
\eea 
Here ${\cal N}_{ij}=N_{m_i-m_j}$ and ${\cal D}_{ij}=\Delta_{m_i-m_j}$ are matrix elements of $M\times M$ matrices. In general, the $M$-kink correlator is a Pfaffian,
\be 
C_{m_1,m_2,\dots, m_M} = 
\left(-1\right)^{\frac12 M(M-1)}
{\rm Pf}
\left( 
\begin{array}{cc}
{\cal D}^\dag  & {\cal N}        \\
  -{\cal N}^T  & {\cal D}
\end{array}
\right),
\label{CM}
\ee 
that can be efficiently calculated numerically. Matrix ${\cal D}$ becomes negligible for sufficient dephasing, and the correlator simplifies to 
\be 
C_{m_1,m_2,\dots, m_M} = {\rm Det}~ {\cal N} .  
\label{CMdephased}
\ee 
For instance, a connected $3$-kink correlator becomes 
$C_{m_1,m_2,m_3}^{(c)}=2 {\cal N}_{12} {\cal N}_{23} {\cal N}_{31}$ in that limit.

\section{ Domain size distribution }
\label{sec:KKdomains}

By definition, the probability that a ferromagnetic domain has size $L$ is:
\be 
P_L =
\frac{1}{\left\langle K_{\frac12} \right\rangle }
\left\langle 
K_{\frac12} \left(1-K_{\frac32}\right) \dots \left(1-K_{L-\frac12}\right) K_{L+\frac12}
\right\rangle .
\label{PLdef}
\ee
This is a conditional probability that, provided that there is a kink on bond $\frac12$, there are no kinks on $L-1$ consecutive bonds $\frac32,\dots,L-\frac12$ and a kink at bond $L+\frac12$. Given that $\left\langle K_{\frac12} \right\rangle=\rho$ and $1-K_{m}=\gamma_{m} \gamma^\dag_{m}$, we can write the distribution as a Pfaffian:
\be 
P_L = 
\rho^{-1}~
{\rm Pf}~ {\cal M} =
\rho^{-1} \sqrt{ {\rm Det} ~ {\cal M} },
\label{PLPf}
\ee 
where ${\cal M}$ is a skew-symmetric matrix of dimensions $2(L+1)\times2(L+1)$. Its elements above the diagonal, for $i<j$, are ${\cal M}_{ij}=\langle \Gamma_i \Gamma_j \rangle$, where operators $\Gamma_1, \dots ,\Gamma_{2(L+1)}$ represent
$
\gamma^\dag_{\frac12}, \gamma_{\frac12}, 
\gamma_{\frac32}, \gamma_{\frac32}^\dag, 
\dots,
\gamma_{L-\frac12}, \gamma^\dag_{L-\frac12},
\gamma_{L+\frac12}^\dag, \gamma_{L+\frac12},
$
respectively. 

Figure~\ref{fig:domains}(a) shows the domain size distribution for several values of the quench time, with both the domain size and the probability rescaled by the KZ length. They do not quite collapse because the anomalous correlators in Eq.~\eqref{Delta}, that contribute to the matrix ${\cal M}$, depend not only on $\hat\xi$ but also on the dephasing length in Eq.~\eqref{l}, which for a linear ramp differs from $\hat\xi$ by a logarithmic correction. However, when a smooth ramp slows in the ferromagnetic phase for long enough to suppress the anomalous correlators, the anomalous correlators can be neglected and the domain size distribution depends on $\hat\xi$ only. The dephased distribution is shown in Fig.~\ref{fig:domains}(a) as a dashed line.

\section{ Emptiness formation probability }
\label{sec:EL}

\begin{figure}[t!]
\vspace{-0cm}
\includegraphics[width=0.99999\columnwidth,clip=true]{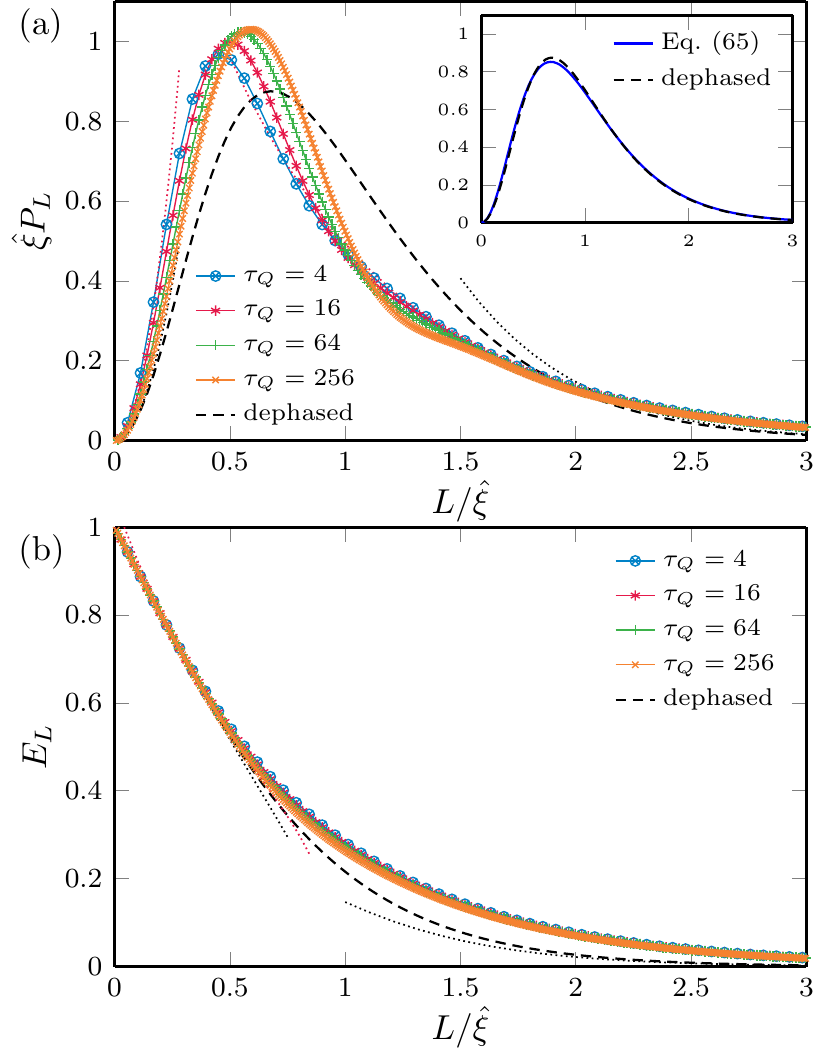}
\vspace{-0cm}
\caption{
{\bf Domain size distribution and emptiness formation probability.}
In the top panel, we show the domain size distribution in Eq.~\eqref{PLPf} for several values of the quench time $\tau_Q$ (solid lines). We present the corresponding EFP in Eq.~\eqref{ELPf} in the bottom panel. The probabilities and size are properly rescaled by the KZ length scale~$\hat\xi$. The scaled plots do not collapse as the curves depend not only on $\hat\xi$ but also on the dephasing length $l$ in Eq.~\eqref{l}.  
The dashed lines show the result of slowing down the ramp in the ferromagnetic phase to allow complete quasiparticle dephasing. The dephasing sends $l\to\infty$, thus making the scaled distribution universal (i.e., independent of $\tau_Q$).
The dotted lines in the top panel are the asymptotes in Eqs.~\eqref{PLas},~\eqref{dephasedPLsmall} (black),~\eqref{PLas2} and~\eqref{PLsmall} (red) for $\tau_Q=16$, and
the asymptotes of EFP in the bottom panel correspond to Eqs.~\eqref{EFPas},~\eqref{dephasedEFPsmall},~\eqref{EFPas2}, and~\eqref{ELsmall}.
The inset of panel (a) compares $\hat \xi P_L$ in the dephased limit (points are data for $\tau_Q=16$), with the formula in Eq.~\eqref{PL_dephased_fit} (blue line).
}
\label{fig:domains}
\end{figure}

The domain size distribution is intimately related to an emptiness formation probability (EFP):
\be 
E_L = 
\left\langle \left(1-K_{\frac12}\right) \dots \left(1-K_{L-\frac12}\right) \right\rangle . 
\label{EL}
\ee 
It is a probability that there are no kinks on $L$ consecutive bonds. Just like the domain size distribution, it is experimentally accessible and, therefore, also interesting in its own right. Given that $1-K_{m}=\gamma_{m} \gamma^\dag_{m}$, the Wick theorem allows one to rewrite EFP as a Pfaffian of a skew-symmetric block Toeplitz matrix:
\be 
E_L = (-1)^{L(L-1)/2}~ {\rm Pf}~ {\cal T} = \sqrt{{\rm Det}~{\cal T}}.
\label{ELPf}
\ee
Here
\be 
{\cal T}=
\left(
\begin{array}{cc}
\tilde{\cal D}    & 1-\tilde {\cal N}           \\
\tilde{\cal N}-1  & \tilde{\cal D}^\dag  
\end{array}
\right).
\label{calT}
\ee
The blocks are $\tilde{\cal D}_{mm'}=\Delta_{m-m'}$ and $\tilde{\cal N}_{mm'}=N_{m-m'}$. 
The EFP has been extensively studied in the Ising/XY chains at equilibrium~\cite{EFP1, EFP2, EFP3,  EFP4}, in which case the block Toeplitz matrics simplifies to a simple Toeplitz matrix (after rotation to the Majorana fermionic operators) due to $\tilde{\cal D}$ being real. The latter is not the case in the state after the linear quench that we consider here and, in general, one has to deal with a full block Toeplitz matrix. However,
with sufficient dephasing, when the anomalous correlators can be neglected, the probability simplifies to
\be 
E_L = {\rm Det}~\left( 1-\tilde{\cal N} \right).
\label{ELDet}
\ee
Figure~\ref{fig:domains}(b) shows $E_L$ in the function of scaled distance, $L/\hat\xi$.

In order to make the connection to $P_L$, we combine Eqs.~\eqref{PLdef} and~\eqref{EL} to find that
\be
\rho P_L  = E_{L+1} + E_{L-1} - 2 E_{L},
\ee
Among others, this implies that a mean domain size is
\be 
\left\langle L \right\rangle \equiv \sum_{L=1}^\infty L P_L = E_0 \hat\xi = \hat\xi,
\label{meanL}
\ee 
where we use the fact that $E_{L}$ is vanishing quickly as $L\to\infty$. Here, $E_0=\left\langle 1 \right\rangle=1$ according to  Eq.~\eqref{EL}. 

Furthermore,  with our general assumption that $\tau_Q\gg1$, or, equivalently, $\hat\xi\gg 1$, the domain size distribution can be brought to a differential form: 
\be 
\hat\xi P_L \approx  
\hat\xi^2 \frac{d^2}{dL^2} E_L.
\label{PLdiff}
\ee 
The usefulness of this equation stems from the fact that, unlike $P_L$, $E_L$ can be defined as a determinant of a Toeplitz matrix, and then $P_L$ can be obtained by its straightforward discrete differentiation. This also opens a route toward some analytic results that we collect in the next section.

Finally, Fig.~\ref{fig:EFPdephasing} shows EFP and domain size distribution when the actual ramp slows down in the ferromagnetic phase as compared to the straight linear ramp in Eq.~\eqref{linear}, see Appendix~\ref{app:slowing}. 
For instance, when near a typical $g_w=1/2$ the ramp waits for a time 
\be
t_w=w~\tau_Q, 
\label{eq:tw}
\ee
then the dephasing length is increased to $l_w$ in Eq.~\eqref{lw}. Increasing $w$ suppresses the anomalous correlator~\eqref{Delta} and drives the distributions towards their dephased limits. To approach the dephased limit around the peak of $P_L$, for relativly fast $\tau_Q=16$ shown in Fig.~\ref{fig:EFPdephasing}(a), the extra waiting time has to be of the order of $10\tau_Q$. The signatures of the anomalous correlator coherence appear to be very robust against the quasiparticle dephasing.

\section{ Asymptotic results }
\label{sec:analytic}

In this section, we derive asymptotes of $E_L$ and $P_L$ valid for either large or small $L$. We begin with the dephased case when the anomalous correlators are negligible.

\subsection{ Dephased limit; asymptotes for large $L$ }

The Toeplitz matrix in Eq.~\eqref{ELDet} has a Fourier transform: 
\be 
\delta_{mn}-N_{m-n} = \frac{1}{2\pi}\int_{-\pi}^{\pi} dk \left( 1-p_k \right) e^{ik(m-n)}
\ee
for $m,n=1,\dots,L$. The generating function has zero at $k=0$, and we factorize it in line with the Fisher-Hartwig conjecture~\cite{schogo1,schogo2,schogo3,schogo4,schogo5},
\be 
1-p_k=1-e^{-(\hat\xi k)^2/4\pi} \equiv (2-2\cos k)^1 ~ \tau(k).
\label{1pk}
\ee 
Then the asymptote for large $L$ reads
\be 
E_L \approx \alpha_D \frac{L}{\hat\xi} e^{-\beta L/\hat\xi},
\label{EFPas}
\ee 
where
\be 
\beta = - \hat\xi \int_{-\pi}^{\pi} \frac{dk}{2\pi}~ \log\tau(k) \simeq 2.6124.
\label{beta}
\ee 
The prefactor in Eq.~\eqref{EFPas} can, in principle, be worked out with the conjecture, but here we treat it as a fitting parameter obtaining $\alpha_D=2.00$. With Eq.~\eqref{PLdiff} we get, in the leading order in $L/\hat\xi$,
\be 
\hat\xi P_L \approx  
\alpha_D\beta^2  \frac{ L}{\hat\xi} e^{-\beta L/\hat\xi}.
\label{PLas}
\ee 
This asymptotes are shown in Fig.~\ref{fig:domains} with black dotted lines.

\begin{figure}[t]
\vspace{-0cm}
\includegraphics[width=0.99999\columnwidth,clip=true]{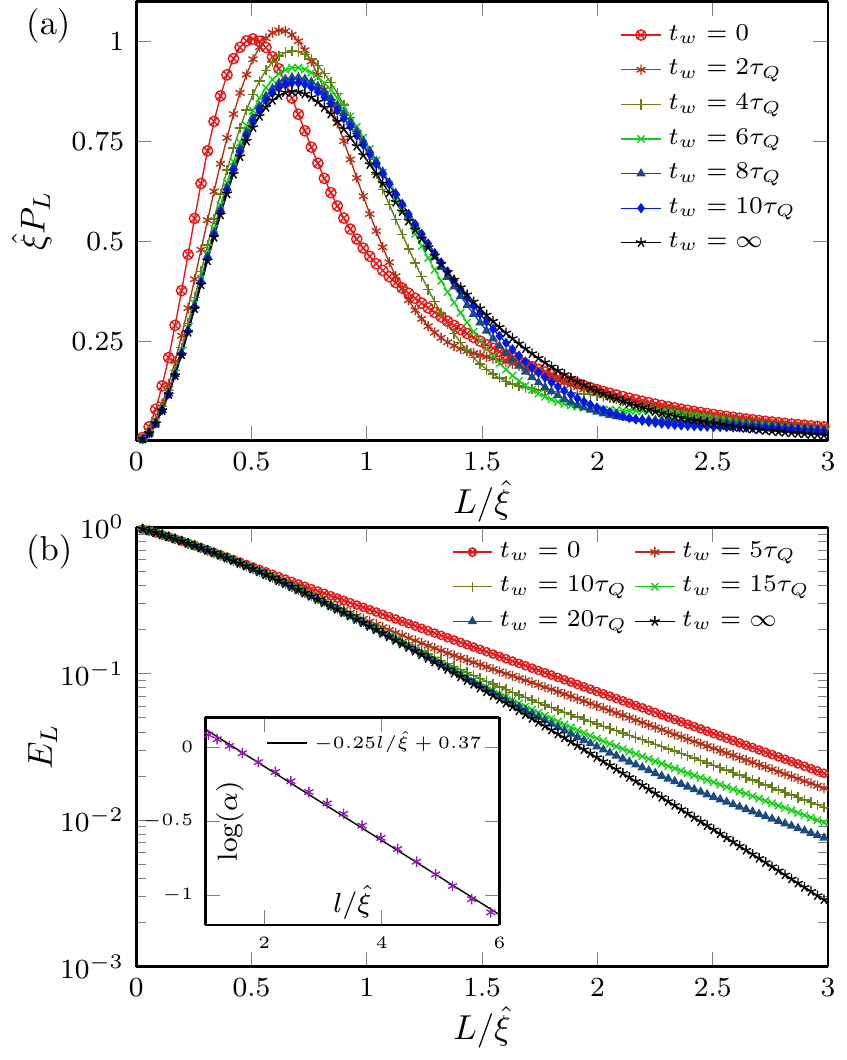}
\vspace{-0cm}
\caption{
{\bf Extra dephasing by slowing the ramp. }
When the ramp slows down near $g=1/2$ in the ferromagnetic phase for an extra waiting time $t_w=w~\tau_Q$, then the anomalous correlator~\eqref{Delta} is suppressed by increasing the dephasing length from $l$ in Eq.~\eqref{l} to $l_w$ in Eq.~\eqref{lw}. 
The top panel shows the domain size distributions in the function of scaled distance for fixed $\tau_Q=16$ and several values of the waiting coefficient $w$. The bottom panel shows the corresponding emptiness formation probability. There is a clear crossover in the asymptotic behavior between coherent and dephased limits with increasing $w$. In the inset, we show prefactor $\alpha$ from Eq.~\eqref{DettildeT} combining data for $t_w/\tau_Q=0,2,4,6,8,10$ for each $\tau_Q=4,16,64$. Those data collapse on a single curve in the function of $l/\hat \xi$.
}
\label{fig:EFPdephasing}
\end{figure}

\subsection{Dephased limit; asymptotes for small $L$}

For small $L$, the Toeplitz matrix in Eq.~\eqref{calT} can be rearranged as a sum of a large and a small matrix:
\bea 
\delta_{mn} - N_{m-n} &=&
\left( \delta_{mn} - \frac{1}{\hat\xi} \right) +
\frac{1}{\hat\xi}
\left( 
1 - e^{-\pi(m-n)^2/\hat\xi^2} 
\right) \nonumber \\
&\equiv&
A_{mn}+B_{mn}.
\label{AB}
\eea 
Matrix $A_{mn}$ has $L-1$ eigenvalues equal to $1$ and one smaller eigenvalue $1-L/\hat\xi$ with eigenvector $\ket{1}=(1,\dots,1)/\sqrt{L}$. Consequently,
\be 
E_L = 
{\rm Det}~(A+B) = 
\left(1-\frac{L}{\hat\xi}\right) ~ {\rm Det}\left(1+A^{-1}B\right).
\ee 
For $L\ll\hat\xi$,  matrix $A^{-1}B$ is small, and we can proceed as
\bea 
&&
\log {\rm Det}\left(1+A^{-1}B\right) = \nonumber \\
&&
{\rm Tr}~ \log \left(1+A^{-1}B\right) \approx 
{\rm Tr}~ A^{-1}B = \nonumber \\
&&
\bra{1} B \ket{1} \frac{L/\hat\xi}{1-L/\hat\xi} \approx 
\frac{L}{\hat\xi}    \bra{1} B \ket{1} .
\eea 
Here, we used ${\rm Tr}~ B=0$. Furthermore,
\bea 
\bra{1} B \ket{1} \approx 
\frac{1}{\hat\xi L} 
\sum_{m,n=1}^L \frac{ \pi (m-n)^2 }{ \hat\xi^2 } \approx 
\frac{ \pi }{ 6 } 
\frac{ L^3 }{ \hat\xi^3 } .
\label{sumsum}
\eea 
Finally, we have 
\be 
E_L\approx\left(1-L/\hat\xi\right)\exp\left(\frac{\pi}{6} \frac{L^4}{\hat\xi^4} \right),
\label{dephasedEFPsmall}
\ee
and with Eq.~\eqref{PLdiff}, in the leading order in $L/\hat\xi$,
\be 
\hat\xi P_L \approx 
2\pi \left(\frac{L}{\hat\xi}\right)^2.
\label{dephasedPLsmall}
\ee 
This asymptotes are shown in Fig.~\ref{fig:domains} with black dotted lines.

\subsection{ Dephased limit;  interpolating formula }

The asymptotes derived in the last two subsections capture the behavior of the tails of $P_L$. However, as can be seen in Fig.~\ref{fig:domains}(a), they diverge from the exact value near the peak of $P_L$. Finding a rigorous expression in this range poses a daunting challenge. Indeed, even including a subleading term in Eq.~\eqref{EFPas} [an additional $\sim 1/L$ term in the exponent~\cite{schogoL}], or the next order term in the expansion of the logarithm in Eq.~\eqref{sumsum},  would not be sufficient to extend the range of applicability of the asymptotic expansions to cover the peak of $P_L$. 

For that reason, we attempt here a simple rational expression that interpolates between the asymptotes in Eqs.~\eqref{PLas} and~\eqref{dephasedPLsmall}:
\be 
\hat\xi P_L =
2\pi r^2 e^{-\beta r} \frac{1+\alpha_D \beta^2 a r}{1+br+2 \pi a r^2}
\label{PL_dephased_fit}
\ee 
Here $r=L/\hat\xi$, $\beta$ and $\alpha_D$ are fixed by the large-$L$ asymptote in Eq.~\eqref{EFPas}. This leaves two parameters, $a$ and $b$, that we adjust by requiring that probability distribution in Eq.~\eqref{PL_dephased_fit} is properly normalized, and the 
mean satisfy the expected condition in Eq.~\eqref{meanL}. 
These two constraints yields: $a=0.3774$ and $b=0.7352$.
We compare this simple interpolation formula with the exact data in the inset of  Fig.~\ref{fig:domains}(a). As can be seen in that plot, it reproduces the exact data quite well despite having no free fitting parameter.

\subsection{ Asymptotes for large $L$ }

The matrix in Eq.~\eqref{calT} can be rearranged as a $L\times L$ Toeplitz matrix of $2 \times 2$ blocks: 
\bea
{\cal T}_{mn} &=&
\left[
\begin{array}{cc}
\Delta_{m-n}           &    \delta_{mn}-N_{m-n}                             \\
N_{n-m}-\delta_{nm}    &    \Delta_{n-m}^*
\end{array}
\right]                                                   \nonumber \\
&\equiv&
\frac{1}{2\pi} \int_{-\pi}^{\pi} dk~ {\cal T}_k ~ e^{ik(m-n)}.
\eea 
Here, the Fourier transform is
\bea
{\cal T}_k &=&
\left[
\begin{array}{cc}
z_k \sqrt{p_k(1-p_k)}    &   ( 1-p_k)              \\
-(1-p_k)                 & z_k^* \sqrt{p_k(1-p_k)}  
\end{array}
\right],
\eea 
where $z_k=\mathrm{sgn}(k)~e^{-i\varphi_k}$.

A central role in determining the asymptotic behavior is played by the determinant of the generating matrix~\cite{widom1}
\be 
{\rm Det} ~ {\cal T}_k = 1-p_k = 1-e^{-(\hat\xi k)^2/4\pi}\equiv (2-2\cos k) ~ \tau(k),
\label{2pk}
\ee
which is identical to the generating function in Eq.~\eqref{1pk} of the simple Toeplitz matrix in the dephased limit. In particular, it does not depend on the dephasing length $l$. The generating determinant has a zero singularity of the Fisher-Hartwig form. While some results on a discontinuous jump singularity are available~\cite{widom_gs1,widom_gs2}, we are not aware of analytical conjecture covering our case. We proceed numerically, obtaining the leading exponential dependence of the form  \footnote{In the numerical calculation of the determinant we discuss here, we used direct numerical integration of the middle line of~\eqref{Delta}, i.e., without the approximation in the third line of that equation. Such approximation, which follows from Eq.~\eqref{Aa}, does not change the functional form in Eq.~\eqref{DettildeT} but would yield slightly different numerical constants. For instance, $\beta$ would be shifted by around $2\%$, consistent with the modification of the geometric mean of the generating function determinant caused by the approximation. }
\be 
{\rm Det} ~ {\cal T} \approx \alpha(l/\hat\xi) e^{-\beta L/\hat\xi},
\label{DettildeT}
\ee 
with $\beta$ given, as expected, by Eq.~\eqref{beta}. We observe that there is no algebraic correction [$\log(L)$ corrections in $\log({\rm Det} ~ {\cal T})$], unlike the ones present in Eq.~\eqref{EFPas}. Dephasing length $l$ appears in a prefactor $\alpha$ that decreases roughly as $\log (\alpha(l/ \hat \xi)) \simeq -\beta_0  l/ \hat \xi + 0.37$, with $\beta_0 = 0.25$, see the inset of Fig.~\ref{fig:EFPdephasing}(b). 

Consequently, the EFP becomes
\be 
E_L \approx \alpha^{1/2} e^{-\frac12\beta L/\hat\xi}
\label{EFPas2}
\ee 
and the domain size distribution follows as
\be 
\hat\xi P_L \approx  
\frac14 \alpha^{1/2} \beta^2 e^{-\frac12\beta L/\hat\xi}.
\label{PLas2}
\ee 
These asymptotes are shown in Fig.~\ref{fig:domains} with red dotted lines.

Interestingly, there is a change in the asymptotic behavior compared to the dephased limit in Eqs.~\eqref{EFPas} and \eqref{PLas}, most notably in the rate of the exponential decay. The emergence of the dephased asymptotics with increasing $l$, resulting from longer waiting time $t_w$, is clearly visible in Fig.~\ref{fig:EFPdephasing}(b). We see that the dephased formulas become valid, roughly, for $L \ll (\beta_0/\beta) l \approx 0.1 l$, while Eqs.~\eqref{EFPas2} and \eqref{PLas2} are relevant for $L \gg 0.1 l$.

\subsection{ Asymptotes for small $L$ }

A more general Toeplitz matrix in Eq.~\eqref{calT} can also be rearranged as a sum of a large and a small matrix, ${\cal T}={\cal A}+{\cal B}$, where
\be 
{\cal A} =
\left(
\begin{array}{cc}
 0 &   A   \\
-A &   0 
\end{array}
\right),~~
{\cal B} =
\left(
\begin{array}{cc}
\tilde{\cal D}  &          B             \\
        -B      & \tilde{\cal D}^\dag 
\end{array}
\right).
\ee 
Here $L\times L$ blocks $A$ and $B$ are the same as in Eq.~\eqref{AB}. Matrix ${\cal A}$ has $(L-1)$ eigenvalues equal to $+i$, $L-1$ equal to $-i$, and two eigenvalues $\lambda_\pm=\pm i (1-L/\hat\xi)$, hence ${\rm Det {\cal A}}=\lambda_+\lambda_-=(1-L/\hat\xi)^2$ and, in accord with Eq.~\eqref{ELPf},
\bea 
E_L
& = & 
\left(1-L/\hat\xi\right)
\exp\left[\frac12{\rm Tr}~\log\left(1+{\cal A}^{-1}{\cal B}\right)\right].
\eea 
With
\be 
{\cal A}^{-1}=
\left(
\begin{array}{cc}
0 & -A^{-1} \\
A^{-1} & 0 
\end{array}
\right),
\ee 
we obtain $\frac12{\rm Tr}{\cal A}^{-1}{\cal B}={\rm Tr}A^{-1}B=\frac16\pi L^4/\hat\xi^4$ and 
\bea
\frac12
{\rm Tr}
\left(
{\cal A}^{-1}{\cal B}
\right)^2 
&=&
{\rm Tr}~\left(A^{-1}B\right)^2-{\rm Tr}~ A^{-1} \tilde{\cal D} A^{-1} \tilde{\cal D}^\dag \nonumber\\
& \approx &
-{\rm Tr}~ \tilde{\cal D} \tilde{\cal D}^\dag 
\approx
-\frac{c^2}{6} \frac{L^4}{\hat\xi l^3}.
\eea
Here we keep only leading order terms in $L/\hat\xi$. Finally, we obtain
\be 
E_L \approx
\left(1-L/\hat\xi\right)
\exp\left[
\frac{L^4}{\hat\xi^4}
\left(
\frac{\pi}{6} + \frac{c^2}{12} \frac{\hat\xi^3}{ l^3}
\right)
\right]
\label{ELsmall}
\ee 
and with Eq.~\eqref{PLdiff} 
\be 
\hat\xi P_L \approx 
\left(
2\pi +
c^2 \frac{\hat\xi^3}{l^3}
\right)
\left(L/\hat\xi\right)^2.
\label{PLsmall}
\ee 
For sufficient dephasing, when $l\gg\hat\xi$, we recover Eq.~\eqref{dephasedPLsmall}.
This asymptotes are shown in Fig.~\ref{fig:domains} with red dotted lines.

\begin{figure}[t!]
\vspace{-0cm}
\includegraphics[width=\columnwidth,clip=true]{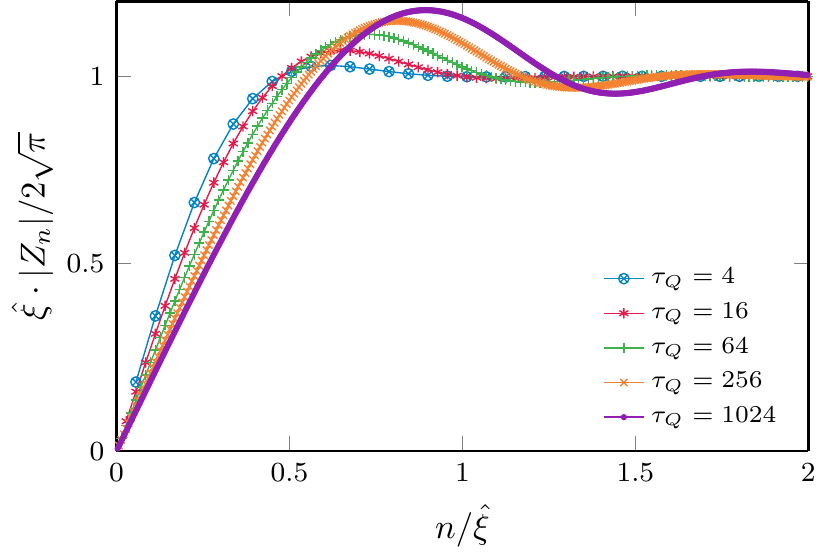}
\vspace{-0cm}
\caption{
{\bf Kink pair wave function. }
The magnitude of the pair wave function in Eq.~\eqref{Zn} scaled by its asymptotic value in Eq.~\eqref{Zinfty} as a function of a scaled distance for several values of quench time $\tau_Q$. The scaled plots do not collapse because the distribution depends not only on $\hat\xi$ but also on the dephasing length $l$ in Eq.~\eqref{l}.  
}
\label{fig:pair}
\end{figure}

\begin{figure}[t!]
\vspace{-0cm}
\includegraphics[width=0.97\columnwidth,clip=true]{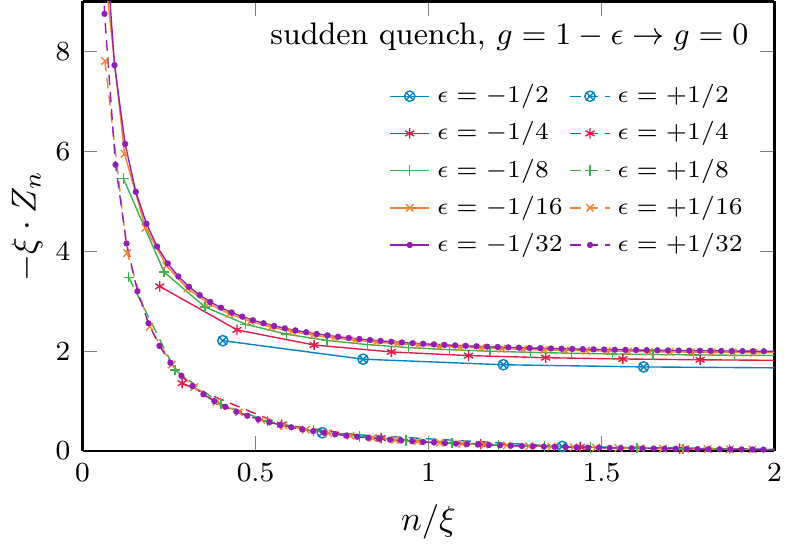}
\vspace{-0cm}
\caption{
{\bf Kink pair wave function after a sudden quench. }
Scaled pair wave function as a function of the scaled distance after a sudden quench from $g=1+\epsilon$ to $g=0$.
For small $\epsilon$, the correlation length in the initial ground state $\xi\propto1/|\epsilon|$. 
After the paramagnetic-to-ferromagnetic quenches, $\epsilon>0$, asymptotes for $n\gg\xi$ are non-zero: $|Z_n|\propto\xi^{-1}$.
The deconfined kinks quenched from the paramagnetic phase destroy the ferromagnetic long-range order.
After the ferromagnetic-to-ferromagnetic quenches, $\epsilon<0$, $|Z_n|$ decays to zero for large $n$ and bounded kink-antikink pairs quenched from the ferromagnetic phase do not affect the long-range order. 
}
\label{fig:pair_quenched}
\end{figure}

\section{ Kink pair wave function }
\label{sec:Cooper}

The Gaussian state is fully characterized by its quadratic correlators, and, as we have seen, they allow efficient computation of a number of experimentally relevant quantities. However, the nature of the fermionic paired state can also be encapsulated in its pair wave function. Although not as computationally potent as the correlators, it still can provide some interpretations. Along these lines, the state in Eq.~\eqref{psi0} can be rewritten as
\bea
\ket{\psi(0)} & \propto & 
\exp\left(
\sum_{k>0}
{\cal Z}_k
\gamma_{-k}^\dag \gamma_k^\dag 
\right)
\ket{\rm GS} \nonumber \\
& = &
\exp\left(
\frac12
\sum_{m,n}
Z_n
\gamma_{m+n}^\dag \gamma_{m}^\dag 
\right)
\ket{\rm GS},
\eea  
with half-integer $m$ in the last line.
Here $\ket{\rm GS}$ is the ferromagnetic ground state~\eqref{GS} without any kinks and 
\be 
{\cal Z}_k=
\sqrt{\frac{p_k}{1-p_k}} 
e^{-i\varphi_k}
\ee 
is a pair wave function. Its position representation is 
\be 
Z_n = -\frac{4}{N} \sum_{k>0} {\cal Z}_k \sin kn.
\ee 
For large $N$, we obtain
\bea 
Z_n & = &
-\frac{2}{\pi} \int_0^\pi dk~ {\cal Z}_k \sin kn .
\label{Zn}
\eea 
For small $k$, ${\cal Z}_k\simeq e^{-i\varphi_0}/\sqrt{2\pi\tau_Q k^2} $, therefore, for large $|n|$ the asymptotes are
\be 
Z_n \approx - 2 \sqrt{\pi} \hat\xi^{-1} ~ e^{-i\varphi_0} ~ {\rm sgn}(n).
\label{Zinfty}
\ee 
This is not a bound kink-antikink pair. The kinks are free to move all along the spin chain except for repulsion when they approach one another, see Fig.~\ref{fig:pair}.

Thanks to the KZ freeze-out at $\hat\epsilon$ above the critical transverse field, this deconfinement is inherited from the paramagnetic phase.  This can be illustrated by a simple example of a sudden quench with $\tau_Q\to 0$.
 In this limit, final $(u_k,v_k)=(1,0)$ at $g=0$ is the same as the initial ground state for $g\to\infty$. Eqs.~\eqref{GS} and~\eqref{ES} then imply ${\cal Z}_k=\cot{k/2}$ and $Z_n=-2~{\rm sgn}(n)$, that indeed has finite asymptotes for large $|n|$.

A more relevant example is a sudden quench from $g=1+\epsilon$ just above the phase transition to $g=0$, see Fig.~\ref{fig:pair_quenched}. The correlation length in the initial ground state is $\xi \propto 1/\epsilon$. The quenched pair wave functions for different $\epsilon$ collapse when scaled with $\xi$. For large $|n|\gg\xi$, they tend to a finite asymptote proportional to density of kinks in the quenched state, $Z_n \propto - \xi^{-1} {\rm sgn}(n)$. Those finite asymptotes demonstrate the deconfinement of kinks quenched from the paramagnetic to the ferromagnetic phase. They are free to move apart all along the chain and destroy the ferromagnetic long-range order.

For comparison, ferromagnetic-to-ferromagnetic quenches from $\epsilon<0$ to $g=0$ result in pair functions that decay quickly to zero when $|n|\gg\xi$, see Fig.~\ref{fig:pair_quenched}. They describe bound kink-antikink pairs that do not affect the long-range order.

Finally, in Fig.~\ref{fig:pair_quenched}, there appear to develop a peak for $\epsilon\to0^\pm$. However, its unscaled magnitude is independent of $\epsilon$, and its maximum is located at $n=1$ (corresponding to individual flipped spins). The peak originates from ultraviolet excitations with large $k$, which are absent in the state after the linear ramp,  reflected by the excitation spectrum in Eq.~\eqref{pk}. This reflects that the freeze-out picture of QKZM is a sudden quench from $+\hat\epsilon$ to $-\hat\epsilon$ instead of all the way down to $g=0$.  

\section{ Conclusion }
\label{sec:summary}

By eliminating some unnecessary approximations~\cite{RadekNowak}, the fermionic kink annihilation operators facilitate calculations of the higher-order kink correlators, the domain size distribution, and the emptiness formation probability. All those quantities are experimentally accessible and sensitive to quantum coherence between eigenstates with different numbers of kinks. The coherences prove to be quite robust to the quasiparticle dephasing in the ferromagnetic phase. Indeed, full dephasing would require deliberately slowing the quench in the ferromagnetic phase for a time equal to several quench times. This robustness makes these observables a convenient test of quantumness for quantum simulators. In fact, the basic kink-kink correlator in Eq.~\eqref{Cc} has already been used as a probe of quantum coherence in Ref.~\cite{King_Dwave1d_2022}, where the quantum Ising chain was realized in the coherent D-Wave quantum annealer. The same Kibble-Zurek quantum quench was also realized in Ref.~\cite{Lukin18} in a chain of Rydberg atoms quantum simulating the Ising chain. In both experiments the state after the quench is measured in the $\sigma^z$ basis. From the outcomes of these measurements, one can infer any correlators obtained in this article. They, in turn, provide insights into the quantum coherence of the experimental set-up. 

\acknowledgments
%
This research was funded by the National Science Centre (NCN), Poland, under project 2021/03/Y/ST2/00184 within the QuantERA II Programme that has received funding from the European Union’s Horizon 2020 research and innovation programme under Grant Agreement No 101017733.
%

\appendix

\section{Anomalous kink correlator}
\label{app:integral}

As in Ref.~\onlinecite{RadekNowak}, we approximate
\bea
\sqrt{p_k(1-p_k)} \approx
e^{-a\pi\tau_Q k^2}
A \sqrt{2 \pi} 
\left(\tau_Qk^2\right)^{1/2},
\label{Aa}
\eea
in order to make the integral in Eq.~\eqref{Delta} analytically tractable. Above, $A$ and $a$ are variational parameters. A convenient choice of $A=19/20$ and $a=4/3$ is very close to the minimum of a Frobenius norm of the difference between the left and the right-hand side. With Eq.~\eqref{Aa}, we obtain
\be 
\Delta_R =
\frac{e^{-i\varphi_0}\sqrt{2}A}{\sqrt{\pi\tau_Q}} 
\int_0^{\infty} q
e^{-a\pi q^2-i q^2 \ln\tau_Q}.
\label{deltabetaapp0}
\ee 
Here $q=\sqrt{\tau_Q}k$ is a scaled pseudomomentum, and the upper limit of the integral is safely extended to infinity for slow enough transitions. The integration yields
\bea 
\Delta_R &=&
c
\frac{R}{\sqrt{\hat\xi l^3}}
e^{ -\frac{3\pi}{2} (R/l)^2 }
e^{-i\phi_R}.
\label{deltabetaapp}
\eea 
Here $c=57\sqrt{6\pi}/80\approx3.0934$ is a constant.

\section{ Dephasing after a non-linear ramp }
\label{app:slowing}

The dynamical phase in Eq.~\eqref{varphik} is relevant for excited modes with small $|k|$ less than $\hat\xi^{-1}\propto\tau_Q^{-1/2}$. This is where $p_k$ in Eq.~\eqref{pk} is non-zero. It arises from the quasiparticle dispersion~\eqref{epsilonk}, which can be expanded for the small $k$ as
\be 
\varepsilon_k \approx 2|1-g| + \frac{g}{|1-g|} k^2.
\label{epsilonkapprox}
\ee 
In the main text we consider a linear ramp $g(t)=t/\tau_Q$ with the time running from $t=-\infty$ to $t=0$. When the linear ramp is a good approximation between $t_c-\hat t$ and $t_c+\hat t$, the spectrum of excitations, $p_k$, remains frozen after $t_c+\hat t$ and does not depend on the actual shape of the subsequent ramp as long as it is smooth. It is not the case for the dynamical phase.

Let us then consider a more general non-linear smooth ramp $\tilde g(t)$, which can be approximated by the linear one within $t_c\pm\hat t$: $\tilde g(t)\approx g(t)$. Without loss of generality we assume $\tilde g(-\tau_Q)=1$ similarly as $g(-\tau_Q)=1$. The non-linear ramp terminates at $t_0$ when $\tilde g(t_0)=0$. When compared to the linear one,  the dynamical phase for the non-linear ramp acquires an extra contribution: 
\bea 
\delta\varphi_k &=& 
2 \int_{-\tau_Q}^{t_0} dt'
\left[ 2|1-\tilde g(t')| + \frac{\tilde g(t')}{|1-\tilde g(t')|} k^2 \right] \nonumber\\
&-&
2 \int_{-\tau_Q}^{0} dt'
\left[ 2|1-g(t')| + \frac{g(t')}{|1-g(t')|} k^2 \right].
\eea 
Given that both the linear and the non-linear ramps depend on time $t'$ through the combination $t'/\tau_Q$, the integral can be worked out as
\be 
\delta\varphi_k =
A~ \tau_Q + B~ \tau_Q k^2.
\ee 
Here $A$ and $B$ are numbers that depend on the non-linear profile of $\tilde g$ in function of $t/\tau_Q$. The dynamical phase in Eq.~\eqref{varphik} gets modified to
\be 
\tilde\varphi_k = \frac{\pi}{4} + \left(2+A\right) \tau_Q + \left( \ln\tau_Q + B \right) k^2\tau_Q.
\ee 
Consequently, the phase and the dephasing length in the anomalous correlator in Eq.~\eqref{Delta} is modified as
\bea
\tilde\phi_R 
&=& 
\frac14\pi+(2+A)\tau_Q-\frac32 {\rm arg}\left(1-\frac{3i\left(\ln\tau_Q+B\right)}{4\pi}\right)+ \nonumber\\
& &
-\frac98(R/\tilde l )^2\left(\ln\tau_Q+B\right), \nonumber \\ 
\tilde l 
&=&
\hat\xi ~ \sqrt{ 1 + \left[\frac{3\left(\ln\tau_Q+B\right)}{4\pi}\right]^2 } .
\label{tildel}
\eea 

In general, a smooth ramp that slows down in the ferromagnetic phase is characterized by $B>0$. Slowing down in the range of $g$`s that are not too small -- keeping away from the flat dispersion $\varepsilon_k=2$ at $g=0$ -- can result in a large value of $B$ that not only increases the dephasing length $\tilde l$, but also suppresses the magnitude of the anomalous correlator in Eq.~\eqref{Delta}. As such, the anomalous correlator gets suppressed by quasiparticle dephasing.

\begin{figure}[t]
\vspace{-0cm}
\includegraphics[width=0.98\columnwidth,clip=true]{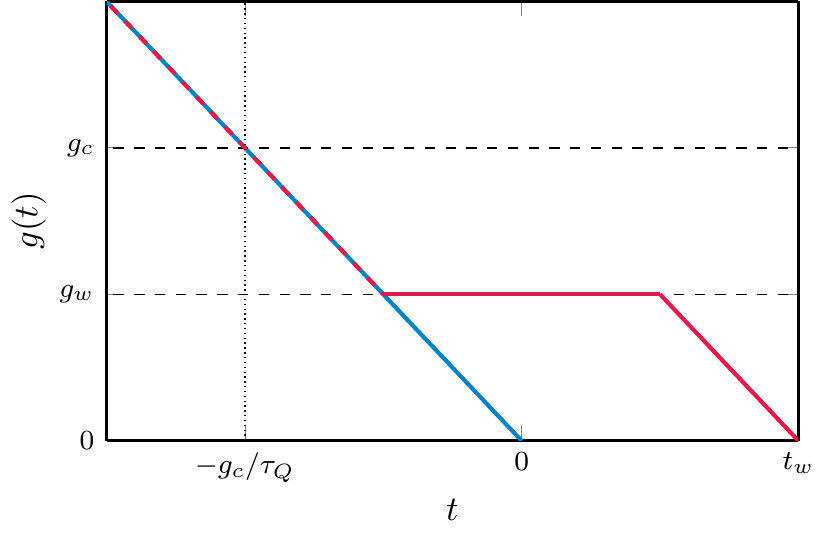}
\vspace{-0cm}
\caption{{\bf Transverse field ramps. }
Time dependence of the transverse magnetic field $g(t)$. The blue line is the standard, straight linear ramp that crosses the quantum critical point at $g_c=1$ and stops at $g(0)=0$ where measurements are performed. The red line is the ramp that stops at $g_w<g_c$ for an extra waiting time $t_w$ before continuing towards $g=0$. The halt allows for an extra quasiparticle dephasing in Eq.~\eqref{deltavarphi}.
}
\label{fig:ramp}
\end{figure}

In order to give a simple example, let us consider a linear ramp that slows at $g_w$, where $0<g_w<1$, for extra waiting time $t_w=w \tau_Q$, before it continues to $g=0$. Its schematic picture is shown in Fig.~\ref{fig:ramp}. For this ramp, the extra dynamical phase is
\be 
\delta\varphi_k = 4(1-g_w) t_w + \frac{2g_w}{1-g_w} t_w k^2,
\label{deltavarphi}
\ee 
or, equivalently, $A=4w (1-g_w)$ and $B=2w g_w/(1-g_w)$. 

For typical $g_w=1/2$, we have $A=2w=B$ and Eq.~\eqref{tildel} becomes
\bea
\phi_{w,R} 
&=& 
\frac14\pi+(2+2w)\tau_Q-\frac32 {\rm arg}\left(1-\frac{3i\left(\ln\tau_Q+2w\right)}{4\pi}\right)+ \nonumber\\
& &
-\frac98(R/l_w)^2\left(\ln\tau_Q+2w\right), \nonumber \\ 
l_w &=& \hat\xi ~ \sqrt{ 1 + \left[\frac{3\left(\ln\tau_Q+2w\right)}{4\pi}\right]^2 }.
\label{lw}
\eea 
The EFP and domain size distributions for several values of $w$ are shown in Fig.~\ref{fig:EFPdephasing}.

%
%

%
\end{document}